\documentclass[noshowpacs,amsmath,
twocolumn,
superscriptaddress,
8pt,
aps,prb]{revtex4-2}
\usepackage[colorlinks=true,citecolor=blue,linkcolor=magenta]{hyperref}
\usepackage{url}
\usepackage{setspace}
\usepackage{amsmath}
\usepackage{multirow}
\usepackage{graphicx}

\usepackage{verbatim}
\usepackage{amsfonts}
\usepackage{amssymb}
\usepackage{epstopdf}
\usepackage{dcolumn}
\usepackage{xcolor}
\usepackage{verbatim}
\usepackage{ulem}
\setcitestyle{super}
\DeclareGraphicsExtensions{.pdf,.eps,.png,.jpg,.mps} 
\begin{document}
\title{Engineered zero-dispersion microcombs using CMOS-ready photonics}

\author{Qing-Xin Ji$^{1,\ast}$, Warren Jin$^{2,4\ast}$, Lue Wu$^{1,\ast}$, Yan Yu$^{1, \ast}$, Zhiquan Yuan$^{1}$, Wei Zhang$^{3}$, Maodong Gao$^{1}$, Bohan Li$^{1}$, Heming Wang$^{1,2}$, Chao Xiang$^{2}$, Joel Guo$^{2}$, Avi Feshali $^{4}$, Mario Paniccia $^{4}$, Vladimir S. Ilchenko$^{3}$, Andrey B. Matsko$^{3}$, John Bowers$^{2,\dagger}$ and Kerry Vahala$^{1,\dagger}$\\
$^1$T. J. Watson Laboratory of Applied Physics, California Institute of Technology, Pasadena, CA 91125, USA\\
$^2$Department of Electrical and Computer Engineering, University of California, Santa Barbara, Santa Barbara, CA 93106, USA\\
$^3$Jet Propulsion Laboratory, California Institute of Technology, 4800 Oak Grove Drive, Pasadena, CA 91109, USA\\
$^4$Anello Photonics, Santa Clara, CA 95054, USA\\
$^{\ast}$These authors contributed equally to this work.\\
$^{\dagger}$Corresponding author: jbowers@ucsb.edu, vahala@caltech.edu}


\maketitle

\noindent\textbf{Abstract}

\noindent Normal group velocity dispersion (GVD) microcombs offer high comb line power and high pumping efficiency compared to bright pulse microcombs. The recent demonstration of normal GVD microcombs using CMOS-foundry-produced microresonators is an important step towards scalable production. However, the chromatic dispersion of CMOS devices is large and impairs generation of broadband microcombs. Here, we report the development of a microresonator in which GVD is reduced due to a couple-ring resonator configuration. Operating in the turnkey self-injection-locking mode, the resonator is hybridly integrated with a semiconductor laser pump to produce high-power-efficiency combs spanning a bandwidth of 9.9 nm (1.22 THz) centered at 1560 nm, corresponding to 62 comb lines. Fast, linear optical sampling of the comb waveform is used to observe the rich set of near-zero GVD comb behaviors, including soliton molecules, switching waves (platicons) and their hybrids. Tuning of the 20 GHz repetition rate by electrical actuation enables servo locking to a microwave reference, which simultaneously stabilizes the comb repetition rate, offset frequency and temporal waveform. This hybridly integrated system could be used in coherent communications or for ultra-stable microwave signal generation by two-point optical frequency division. 

\medskip

\noindent\textbf{Introduction}

\begin{figure*}[ht!]
\includegraphics[width=\linewidth]{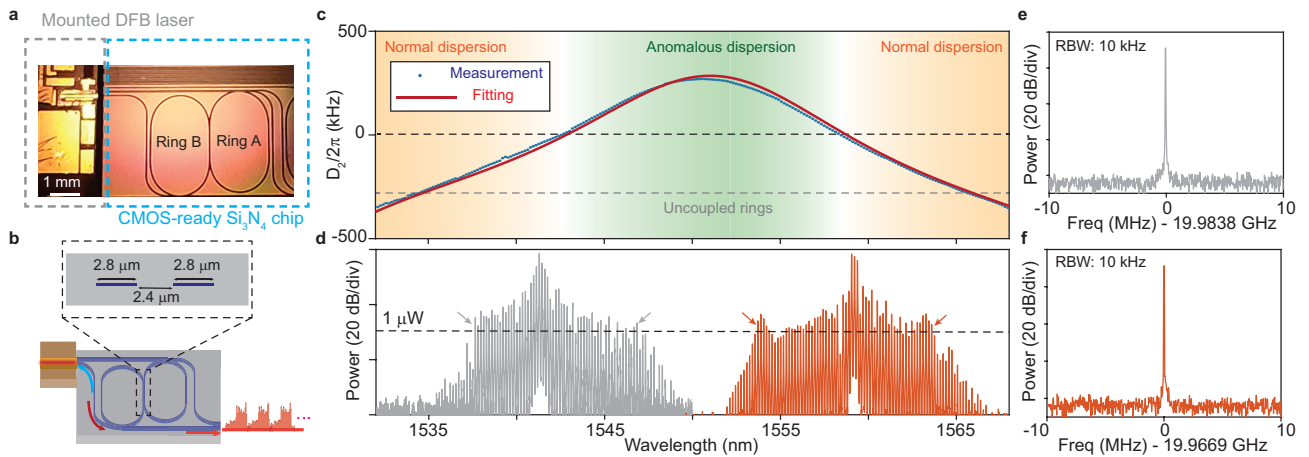}
\caption{{\bf Self-injection-locked zero-dispersion microcomb using a coupled-ring resonator. } 
{\bf a,} Photograph of the coupled ring resonators showing DFB pump laser. The two rings are nearly degenerate ($FSR$ difference is $\sim 100$ MHz). 
{\bf b,} Lower panel shows the coupled rings pumped by a DFB laser which is self-injection locked to a resonator mode. Periodic pulse trains (light red arrow) are generated. Upper panel shows the cross-sectional waveguide geometry of the ring coupling region (2.8 $\mu$m wavegudie width and  2.4 $\mu$m waveguide gap). 
{\bf c,} Measured group velocity dispersion (GVD) parameter $D_2$ of the mode resonances versus wavelength. Zero dispersion is achieved near $1542$ nm and $1559$ nm. 
{\bf d, } Optical spectra of the generated frequency comb when pumped near the two zero dispersion points. For the comb pumped at 1541 nm (gray), the span is 9.2 nm (as marked by the arrows). For the comb pumped at 1559 nm (orange), the span is 9.9 nm. The dashed line indicates the estimated on-chip line power of 1 $\mu$W. 
{\bf e, f, } Measured microwave frequency spectrum of the detected microcomb output when pumped at 1542nm (e) and 1560 nm (f). The resolution bandwidth is 10 kHz. 
}
\label{figure1}
\end{figure*}

\begin{figure*}[!ht]
\centering
\includegraphics[width=\linewidth]{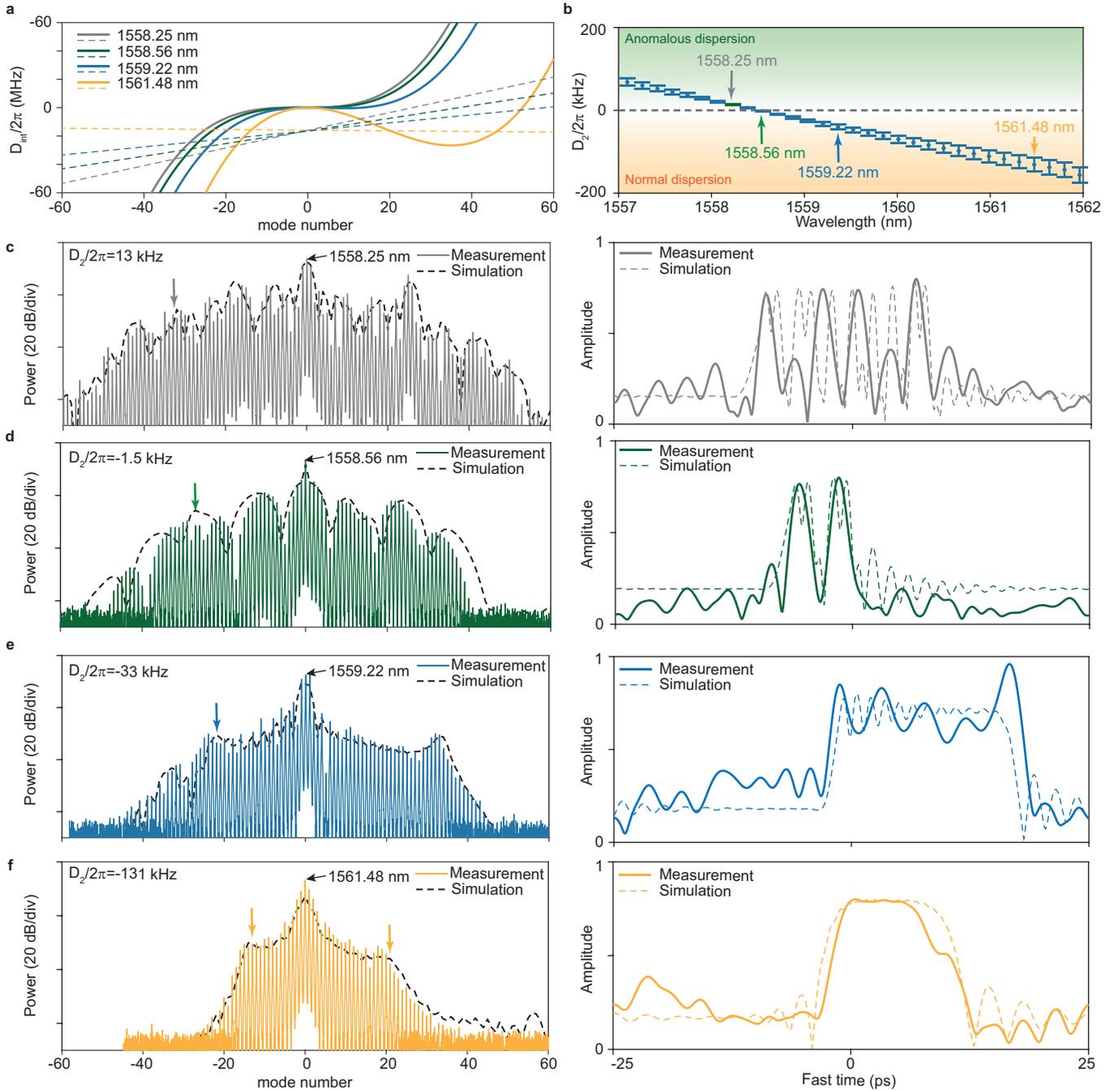}
\caption{{\bf Microcomb optical spectra and temporal waveforms measured at specific pumping wavelengths near the zero GVD wavelength.} 
{\bf a,} Solid lines give the integrated dispersion, defined as $D_{\rm int}=\omega_{\mu}-\omega_0-D_1 \mu$ (right-hand side of Eqn. \ref{eqn:dispersive_wave}), measured at pumping wavelengths shown in the legend. The colored dashed lines denote the comb line frequencies (left-hand side of Eqn. \ref{eqn:dispersive_wave}) when pumped at the specified pumping wavelengths. The intersection of the mode resonances (solid line) and comb lines (dashed line) generates a dispersive wave. 
{\bf b,} Measured GVD parameter $D_2$ versus wavelength near the zero GVD point. Arrows correspond to the pump wavelengths shown in panel a. Errorbar denotes standard deviation from the third-order polynomial fitting. 
{\bf c-f,} Left panels: optical spectra of the microcomb when pumped near the zero-dispersion wavelength (specific pumping wavelength indicated and correspond to values in panel a). Simulated optical spectrum is plotted as the dashed black curve. The dispersive wave position from panel a is marked by arrows. 
Right panel: measured microcomb temporal waveform during one round trip for the pumping wavelength given in left panel.
}
\label{figure2}
\end{figure*}

\begin{figure*}[t]
\includegraphics[width=\linewidth]{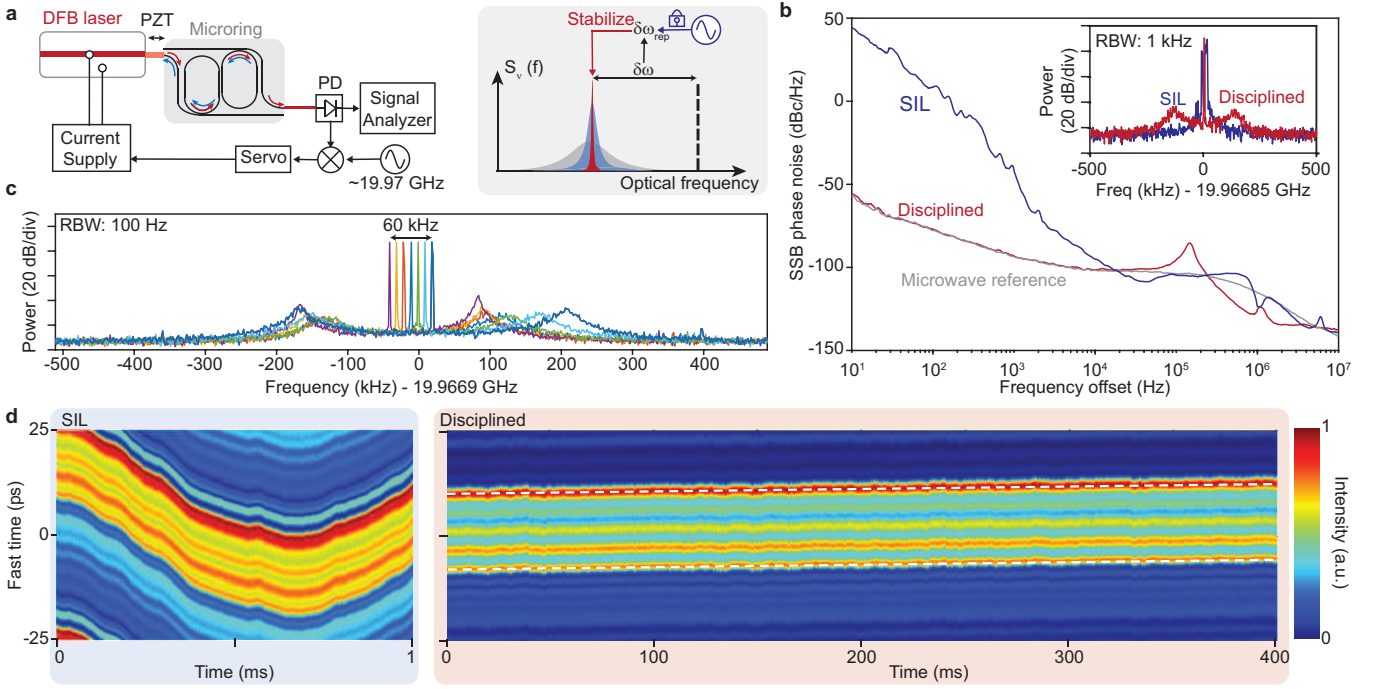}
\caption{{\bf Discipline of the SIL microcomb repetition rate to an external microwave reference. } 
{\bf a,} Experimental setup for external disciplining of the SIL microcomb repetition rate. 
PZT: Piezo-electric transducer. PD: photo detector. Servo: laser servo. Inset: schematic view of laser detuning $\delta\omega$ stabilization via repetition rate discipline. Fluctuation in $\delta\omega$ leads to fluctuation in repetition rate $\delta\omega_{\rm rep}$, which is measured and compared to a microwave reference to generate the error signal for feedback to the laser driving current.  Frequency noise of the free-running DFB laser (gray) is reduced by SIL (blue) and then reduced further by repetition rate locking (red). 
{\bf b,} Repetition rate phase noise and microwave spectra of the free-running SIL microcomb (blue) and the externally disciplined microcomb (red). Measured phase noise of the microwave reference is plotted in gray. 
{\bf c,} Repetition rate tuning of the external disciplined microcomb. Measured microwave tone of the comb is plotted in different colors. The resolution bandwidth is 100 Hz. 
{\bf d, } Stability of temporal waveform of the microcomb under free-running SIL (left) and external discipline (right). In the right panel, two dashed white lines give the linear motion of the waveform in the fast frame. 
}
\label{figure3}
\end{figure*}

\begin{figure}[!ht]
\includegraphics[width=\linewidth]{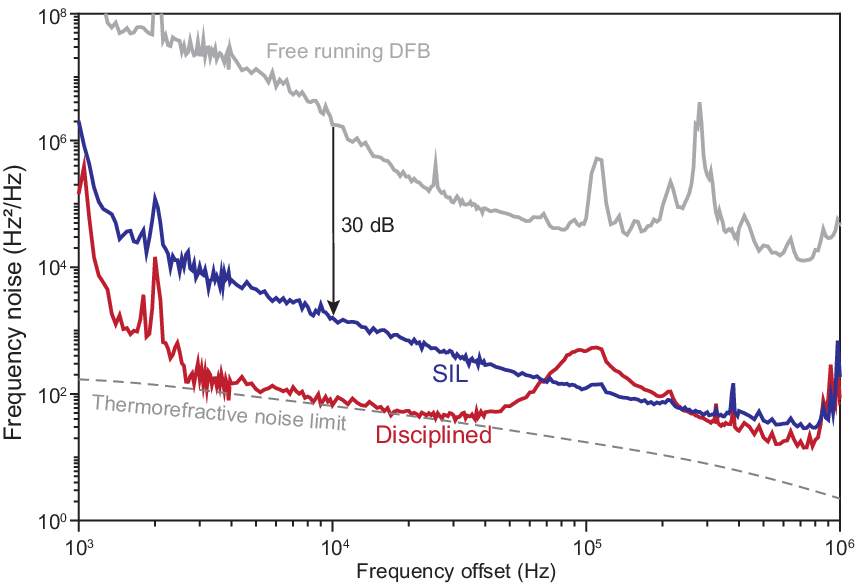}
\caption{{\bf Optical frequency noise of the pump laser (comb offset frequency). } 
Grey curve gives the free-running (non SIL) measurement, blue curve is for self injection locked (SIL) operation, and the red curve is for repetition rate discipline using an external microwave reference. The thermorefractive noise floor is denoted by the dashed gray line.
}
\label{figure4}
\end{figure}

Soliton mode locking in optical microresonators \cite{vahala2003optical} is receiving intense interest for chip-scale integration of frequency comb systems \cite{Kippenberg2018,diddams2020optical,chang2022integrated}. An important advancement has been the realization of microcombs that are directly pumped by semiconductor lasers without amplification \cite{stern2018battery,shen2020integrated,jin2021hertz,xiang2021laser}. These systems have resulted from steady progress in boosting of resonator $Q$ factor so as to lower pumping power, especially in detectable-rate microcombs \cite{herr2014mode,yi2015soliton,liang2015high,suh2018gigahertz,yang2018bridging,liu2020photonic,jin2021hertz}. The directly pumped sytems benefit from self-injection-locking (SIL) of the pump by feedback from the microcomb resonator \cite{kondratiev2017self, pavlov2017soliton}. SIL operation simplifies integration by eliminating the optical isolator component between pump and microcomb, and it also narrows the pump line. Critically, it has also been shown that SIL tends to simplify the soliton turn-on process, making it deterministic (or turnkey) for bright solitons \cite{shen2020integrated}. 

Normal GVD dispersion microcombs  \cite{xue2015mode} have also been shown to benefit from SIL operation \cite{jin2021hertz,wang2022self,lihachev2022platicon}. Not only does the pulse triggering become deterministic, but switching-wave stability dynamics that normally favor large or very small duty cycles, are overcome and pulse duty cycle lies closer to the ideal 50$\%$ for maximal comb power and efficiency\cite{wang2022self}. This is advantageous for microwave generation as well as for use of the microcomb as a WDM communications source \cite{fulop2018high}. Despite these useful properties of the normal GVD SIL microcombs, the spectral width of these systems is limited. For example, in high-$Q$ CMOS-ready resonators, comb lines extend to about 4 nm due to the strong normal chromatic dispersion of the low-confinement waveguides \cite{jin2021hertz,xiang2022silicon}. An intriguing approach to extend bandwidth has been to drive microcombs near the zero GVD wavelength where pulse formation is influenced by higher-order dispersion \cite{anderson2022zero,zhang2022microresonator,li2021observations}. Besides permitting a range of new pulse behaviors as the system is operated above, below, or near the zero-dispersion wavelength, flattening in the dispersion spectrum is generally possible, which broadens the comb span. Under pulsed pumped operation near octave span detectable repetition rates have been possible \cite{anderson2022zero}. 

Here, dual coupled-ring (CR) resonators are used to produce near-zero-GVD microcombs using the normal dispersion CMOS-ready platform. While prior zero-GVD systems have used pulse pumping or optical amplification of continuous-wave sources, the CR resonators feature high intrinsic $Q$ factors over 100 million (the highest of any CR resonator system), enabling SIL operation of the microcombs with a heterogeneously-integrated III-V laser. This is also the first application of the SIL mode of operation to zero-GVD on-chip systems and it is observed to provide the turnkey benefit previously observed for anomalous and normal dispersion systems, including high efficiency comb operation \cite{wang2022self}. Comb bandwidth up to 9.9 nm (1.22 THz) is set by design for application in a two-point optical frequency division system \cite{papp2014microresonator,li2014electro}. This significantly improves upon prior CMOS-ready designs and also offers high comb line powers required in the division system. 
To our knowledge, the number of combs lines generated is a record for non-amplified and non-pulse-pumped zero-GVD systems. The precise stepper-lithography-defined geometry of the CMOS-ready system also enables good control of the zero-GVD wavelength. By tuning the pumping laser near the zero-dispersion point, various mode-locked comb states at microwave rates are observed. 
And by utilizing the linear optical sampling technique assisted with an electro-optic frequency comb \cite{yi2018imaging}, the temporal profiles of these states are measured for the first time including soliton molecules \cite{anderson2022zero,zhang2022microresonator}, switching waves \cite{lihachev2022platicon,wang2022self} and their hybrids \cite{anderson2022zero}. 
Numerical modeling is in agreement with the optical spectra as well as the measured temporal profiles. Servo locking of the microcomb repetition rate to a microwave reference is also demonstrated and stabilized the comb repetition rate, offset frequency and temporal waveform.

\section{High-$Q$ CMOS coupled ring system}

In the experiment, high-$Q$ silicon nitride (SiN) coupled rings are fabricated using the CMOS-ready process \cite{jin2021hertz} (Fig. \ref{figure1}a). The silicon nitride layer thickness is chosen as 100 nm to maintain single mode propagation while also boosting $Q$ factor \cite{jin2021hertz}. Intrinsic $Q$ factors over 100 million are achieved (Supplementary Note 1). The resonators are overcoupled with two bus waveguides (Fig. \ref{figure1}a,b) to enhance comb efficiency (loaded $Q$ factor is $26$ million). The uncoupled resonators feature normal dispersion and will generate dark pulse combs when operated in the SIL configuration. Strong resonator coupling is introduced to lower the dispersion, creating two zero GVD wavelengths (around 1542 and 1559 nm) as shown in Fig. \ref{figure1}c. A discussion on repeatability of dispersion between different devices is included in Supplementary Note 5. 

Coupled ring resonators \cite{boeck2010series} offer a convenient way to engineer anomalous dispersion into systems comprised of normal dispersion waveguides \cite{soltani2016enabling,kim2017dispersion}. And recently they have been studied in the context of normal GVD solitons to engineer either zero-GVD \cite{anderson2022zero} or controlled (as opposed to accidental \cite{xue2015mode}) avoided mode crossing through hybridization of a single mode pair in a photonic molecule \cite{helgason2021dissipative}. Here, this hybridization is engineered to be broadband so as to spectrally flatten the dispersion curve and make higher-order dispersion important. This is done by first introducing strong mutual coupling of rings wherein nearly 40\% of mode power is coupled between resonators, and second by close matching of the ring free-spectral range (FSR) values (the FSR difference is 100 MHz on FSRs close to 20 GHz). The $Q$ factors of the coupled rings are also much higher, enabling both detectable-rate comb operation and hybrid integration with a III-V pump.

\section{Dispersion and SIL microcomb spectra}

In the measurement, a DFB laser operating in the vicinity of one of the two zero-dispersion wavelengths is butt-coupled to the resonator with $\sim 30-40$ mW power coupled onto the waveguide (Fig. \ref{figure1}a,b). Rayleigh scattering inside the resonator reflects $\sim 2\%$ of the power into the pumping laser. 
An Emcore Corp. DFB laser is used for pumping near 1559 nm and a PhotonX Inc. DFB laser is used for pumping near 1541 nm. Temperature tuning of the DFB lasers allows fine tuning control of the pumping wavelength for access to slightly anomalous, near-zero, and slightly normal dispersion wavelengths of the resonator. Fig. \ref{figure2}a shows measurements of integrated dispersion of the resonator for pumping at specific wavelengths (provided in the legend) near 1559 nm. In the plots, $\mu = 0$ corresponds to the pump line. The dispersion curves are measured using a radio-frequency calibrated interferometer reference \cite{yi2015soliton}.

When pumped at near-zero dispersion, the comb spectral span is strongly influenced by comb radiation into a dispersive wave, which corresponds to cavity modes that are nearly resonant with comb lines. 
The mode number $\mu_{\rm dw}$ of the dispersive wave is given as the solution to the equation \cite{li2020experimental},
\begin{equation}
    -\delta\omega+2 g P_0 + \delta\omega_{\rm rep} \mu_{\rm dw}
    =\frac{D_2}{2}\mu_{\rm dw}^2+\frac{D_3}{6} \mu_{\rm dw}^3, 
    \label{eqn:dispersive_wave}
\end{equation}
where $\delta \omega$ is the pump-laser cavity detuning as regulated by the self-injection feedback, $g$ is Kerr nonlinear coefficient, $P_0$ is photon number of the pumped cavity resonance, $\delta\omega_{\rm rep}$ is the difference between comb repetition rate and cavity free-spectral range ($D_1$), and $D_2$ and $D_{3}$ are second- and third-order dispersion parameters, respectively. Fig. \ref{figure2}b gives the measured second-order dispersion ($D_2$) parameter at these same pumping wavelengths, and $D_{3}=7.5$ kHz is used for all pumping wavelengths. The left side of Eqn. \ref{eqn:dispersive_wave} corresponds to the frequency comb lines and is plotted as the colored dashed lines in Fig. \ref{figure2}a. In making these these line plots, $\delta\omega$ and $\delta\omega_{\rm rep}$ are numerically simulated (Supplementary Note 4). The right side of Eqn. \ref{eqn:dispersive_wave} is the integrated dispersion and fitted to the experimentally measured dispersion as noted above. At each pumping wavelength, the dispersive wave mode number $\mu_{\rm dw}$ can be obtained as the intercept between the comb frequencies (dashed lines) and the corresponding integrated dispersion curve (solid lines). 

The dispersive wave position is marked with an arrow in the left panel of Fig. \ref{figure2} c-f, which show measured spectra for SIL pumping at the wavelengths in Fig. \ref{figure2}a. The arrow position provides a trend of measured comb span for negative mode numbers. Overall comb spectral span tends to be determined by this dispersive wave.  Note that in Fig. \ref{figure2}, the third dispersive wave (after mode number $\sim$50) is absent in both the measured and simulated optical spectrum, which is a result of insufficient pump power. The comb spectrum in Fig. \ref{figure1}d spans 9.9 nm (1.22 THz) and features on-chip comb-line power higher than 1 $\mu$W, marked with arrows. This comb has superior spectral coverage compared to previous SIL operated normal GVD microcombs \cite{jin2021hertz,lihachev2022platicon}. It also features an increased number of comb lines compared with higher repetition rate non-bright combs directly III-V pumped with optical isolation \cite{shu2021sub}. When pumped at 1541 nm near the other zero-dispersion wavelength, a microcomb spanning 9.2 nm (1.16 THz) is realized. The comb states both feature pump power conversion efficiency as high as 26\% (see Supplementary Note 2). These comb spectra are plotted in Fig. \ref{figure1}d.

\section{Imaging of mode-locked microcomb states }

While the temporal envelope of pulses produced by normal GVD combs has been measured using cross-correlation \cite{xue2015mode}, near-zero-GVD operation allows access to a wider variety of comb states including soliton molecules (bounded bright solitons) and switching waves \cite{garbin2017experimental,li2020experimental}. These interesting states have been analyzed numerically \cite{parra2014third,parra2017coexistence,anderson2022zero}, but they have so far not been observed in the time domain. This is result of current zero-dispersion systems being generated with a pulsed pump where the comb waveform is influenced by the pump pulse waveform or at challenging high repetition rates \cite{anderson2016measurement,zhang2022microresonator,xiao2021zero}. Here, the linear optical sampling technique assisted with an electro-optic frequency comb is implemented to image the temporal profile of the various comb states \cite{yi2018imaging,wang2022self}. The sampling EO comb spans $\sim 5$ nm with 33 lines, corresponding to a $\sim 1.5$ ps pulse in the time domain, and its repetition rate is set to be slightly higher than the zero-dispersion microcomb. Note that the temporal duration of the soliton pulse (for example, 2.6 
 ps FWHM in Fig.\ref{figure2}d for the left pulse in measurement) is close to the pulse width of the EO-comb ($\sim$ 1.5 ps), thus the imaging result may not perfectly resolve fine structure in the temporal waveforms. By combining the EO comb and the generated comb at the ``drop'' port, and detecting with a fast photodetector, the sampled microcomb signal is recorded by an oscilloscope for processing. 

Experimental results are presented in the right panels of Fig, \ref{figure2}c,d,e,f. At the anomalous side ($D_2>0$) of the near-zero GVD wavelength, the strong dispersive wave binds several solitons into soliton clusters, as observed in Fig.\ref{figure2}c (right panel). The interference between the pulses creates the multiple fringes that are apparent in the optical spectrum (left panel). Numerically simulated time domain and spectra are also presented as dashed curves in Fig, \ref{figure2}c,d,e,f. The clustered soliton formation is preserved for near-vanishing GVD ($D_2/2\pi=-1.5$ kHz) as shown in Fig, \ref{figure2}d, where a  soliton dimer is imaged.  The corresponding optical spectrum is shown in the left panel of Fig. \ref{figure2}d. For the pumping wavelength with normal dispersion, the resulting waveforms are often called switching waves or ``platicons'' \cite{lihachev2022platicon,wang2022self}. Here, the waveform evolves towards a square pulse as the dispersion becomes more normal ($D_2<0$) (Fig. \ref{figure2}e-f, right panels). The present results experimentally demonstrate the evolution of comb states in the near zero GVD regime with different GVD sign, and are consistent with a previous numerical study \cite{anderson2022zero}. 

Generally, operation of the comb with a small amount of normal dispersion (e.g. the state in Fig. \ref{figure2}e, pumped at 1559.2 nm) provides both good spectral coverage as well as a temporal and spectral waveform that is more regular in shape. Furthermore, the square pulse nature of these states (apparent in Fig. \ref{figure2}e-f, right panels) offers an increased duty cycle, which boosts comb power conversion efficiency. As a result, the microwave and optical performance of the comb state in Fig. \ref{figure2}e is further studied below. 

\section{Discipline of the comb repetition rate} 

Discipline of comb repetition rate to an external reference such as a clock is important in many comb applications \cite{diddams2020optical}. Generally, the repetition rate of microcombs is regulated by the pump-laser cavity detuning $\delta\omega$ \cite{yi2017single,lucas2020ultralow}, via channels including dispersive wave recoil \cite{yi2017single}, Raman self-frequency shift \cite{yi2016theory} and inhomogeneous back scattering \cite{wang2022self,kondratiev2020numerical}. And with laser cavity detuning $\delta\omega$ controlled by the applied current on the pumping DFB laser, a fast feedback loop can be used to stabilize the microcomb's repetition rate.

In the experiment (Fig. \ref{figure3}a), the microcomb's repetition rate is detected at the resonator drop port by a fast photo detector, and analyzed by a signal analyzer (R\&S FSUP). Pumping is at 1559.22 nm. The detected repetition rate tone is simultaneously split by a directional coupler after electrical amplification and mixed with a local oscillator which serves as the reference (R\&S SMB 100A). The mixed-down signal is sent to a servo controller and fed back to the current supply (LDX-3620B, DC-coupled modulation response bandwidth $<$1 MHz) of the DFB laser to provide fine tuning control of the pump-laser cavity detuning frequency. The gap distance between the bus waveguide facet and the DFB laser head is regulated by a closed loop piezo (PZT) with a built-in strain gauge displacement sensor (Thorlabs MAX311D).

The measured phase noise of the detected repetition rate tone under open-loop and disciplined (locked) conditions is shown in Fig. \ref{figure3}b. The phase noise of the microwave reference is shown in gray. The phase noise of the disciplined repetition rate follows that of the microwave reference within the feedback bandwidth of approximately $100$ kHz (as defined by the offset frequency where a ``bump'' is observed in its frequency spectrum). The microwave spectra of the SIL and disciplined comb are plotted in inset of Fig. \ref{figure3}b. The repetition rate could be tuned by 60 kHz through tuning of the microwave reference (Fig. \ref{figure3}c). Disciplining the microcomb simultaneously stabilizes its temporal waveform. This is observable in Fig. \ref{figure3}d which shows round trip comb field plotted versus time. In the left panel the data are presented under unlocked conditions and show drift of the soliton waveform over a time scale of 1 ms. On the other hand, the disciplined microcomb features a measured soliton waveform that evolves very slowly and linearly (linear trend lines shown as white dashed lines). Here, the observable drift results from  the difference in sampling period between the oscilloscope and EO sampling signal. All of the sampling data presented in Fig. \ref{figure2} were obtained when the comb was externally disciplined.

\section{Comb offset frequency stabilization under repetition rate discipline} 

The pump laser frequency of the microcomb also determines the offset frequency of the microcomb. And self-injection locking (here of the pump laser) is known to improve frequency stability \cite{liang2010whispering}. In the present case, SIL stabilizes the DFB pump laser's frequency to the high-$Q$ silicon nitride cavity mode and thereby stabilizes laser-cavity detuning ${\delta \omega}$ to a value that depends on the feedback phase and initial laser-cavity detuning without SIL feedback. The pump laser frequency noise under SIL and free-running operation is measured in Fig. \ref{figure4}. The measurement is performed by amplifying the microcomb output at the drop port with an erbium-dope amplifier (EDFA) and filtering out the pump line with a waveshaper. Frequency noise of the line is then measured with a cross-correlation-based self-heterodyne measurement \cite{yuan2022correlated} within the measurement time of 400 ms. Compared with the free running DFB laser (gray), frequency noise of the pump laser is reduced by 30 dB under SIL operation (blue) at 10 kHz offset frequency.
An integrated linewidth $\Delta \nu_{\rm c}$ of 7.2 kHz is obtained ($\Delta \nu_{\rm c}\equiv \int_{\Delta \nu_c}^{+\infty} S_{\delta\omega}(f)/f^2 \mathrm{d}f=\frac{1}{2\pi}$). The frequency noise here is still higher than the thermal noise limit given by the dashed gray curve in Fig. \ref{figure4} \cite{kondratiev2018thermorefractive} (see Supplementary Note 1).

Fluctuation of a microcomb's repetition rate $\delta\omega_{\rm rep}$ is largely controlled by the pump laser-cavity detuning $ {\delta\omega}$ \cite{lihachev2022platicon,kondratiev2020numerical}. As schematically illustrated in the inset of Fig. \ref{figure3}a, disciplining the microcomb's repetition rate thus simultaneously disciplines the laser pumping frequency frequency relative to the pumped cavity resonance. 
With fluctuation in ${\delta \omega}$ greatly reduced by disciplining the repetition rate, frequency noise of the pump laser frequency (and comb offset frequency) is further stabilized to the thermo-refractive noise limit, as plotted in red in Fig. \ref{figure4}. The bumps at 1 kHz and 2 kHz originate from the piezo-electric motor used to control the gap between the DFB laser head and the photonic chip. In principle, heterogeneous integration or butterfly packaging\cite{shen2020integrated} of the pump and the resonator would eliminate the need for this servo control or at least replace it with a potentially faster equivalent control.
It is noted that comb offset noise at lower offset frequencies than the measurement in Fig.\ref{figure4} is generally possible from (but not limited to) the ambient including temperature fluctuations and mechanical vibrations, other than the noise sources noted above \cite{newbury2007low}. 
With the same definition as above, the calculated integrated linewidth is 2.1 kHz, which is comparable to an ECDL pumped microcomb \cite{lei2022optical}. 

\section{Summary}

We have demonstrated a microwave-rate mode-locked microcomb near zero GVD using a CMOS-ready coupled ring resonators. The reduced GVD enables broader microcomb operation using integrated photonics with high conversion efficiency, while maintaining the benefits of the self-injection feedback, including turnkey operation and optical linewidth reduction. A record number of comb lines is generated for non-amplified and non-pulse-pumped operation of a normal dispersion microcomb. The dispersion engineering scheme in this paper can be extended to other wavelengths and different photonic platforms. Disciplining the microcomb to an external microwave reference simultaneously stabilizes its microwave tone, temporal waveform and optical frequency. The external discipline for the hybrid-integrated microcomb source demonstrated here can be used in a two-point optical frequency division system \cite{li2014electro} for low-noise microwave synthesis. 

\bigskip


\bibliography{ref.bib}

\begin{footnotesize}
\noindent 
\end{footnotesize}

\noindent\textbf{Funding.}  Defense Advanced Research Project Agency GRYPHON (HR0011-22-2-0009) and APHI (FA9453-19-C-0029) programs; NASA.
\medskip

\noindent\textbf{Acknowledgments.} The authors thank H. Blauvelt at EMCORE Corporation for supplying the DFB laser used in this study, as well as N. Kondratiev and V. Lobanov for discussions on numerical modelings. 
The authors also thank S. Diddams and F. Quinlan for fruitful comments on the results. The reported here research performed by W. Z., V.S.I. and A.B.M. was carried out at the Jet Propulsion Laboratory, California Institute of Technology, under a contract with the National Aeronautics and Space Administration (80NM0018D0004).
\medskip

\noindent\textbf{Disclosures.} The authors declare no conflicts of interest.
\medskip

\newpage

\end{document}